**Assessing the Value of Complex Refractive Index and Particle Density for Calibration of Low-Cost Particle Matter Sensor for Size-Resolved Particle Count and PM2.5 Measurements**

Ching-Hsuan Huang [a], Jiayang He [b], Elena Austin [a], Edmund Seto [a], Igor Novosselov [b,*]

[a] Department of Environmental and Occupational Health Sciences, School of Public Health, University of Washington, Seattle, WA 98115, United States

[b] Department of Mechanical Engineering, College of Engineering, University of Washington, Seattle, WA 98115, United States

* Corresponding author.

E-mail address: ivn@uw.edu

**Abstract**

Commercially available low-cost PM sensors provide output as total or size-specific particle counts and mass concentrations. These quantities are not measured directly but are estimated by the original equipment manufacturers' (OEM) proprietary algorithms and have inherent limitations because particle scattering depends on the particles' composition, size, shape, and complex index of refraction (CRI). Therefore, there is a need to characterize and calibrate their performance under a controlled environment. Here, we present calibration algorithms for Plantower PMS A003 particulate matter sensor as a function of particle size and concentration. A standardized experimental protocol was used to control the PM concentration, environmental conditions and to evaluate sensor-to-sensor reproducibility. The calibration was based on tests when PMS A003 sensors were exposed to different polydisperse standardized testing aerosols. The results suggested particle size distribution from PMS A003 sensor was shifted compared to reference instrument measures. For calibration of number concentration, linear model without adjusting aerosol properties, including CRI and relative humidity, corrects the raw PMS A003 sensor measurement for specific size bin with normalized mean absolute error within 4.0% of the reference instrument. Although the Bayesian Information Criterion (BIC) suggests that models adjusting for particle optical properties and relative humidity are technically superior, they should be used with caution as the particle properties used in fitting were within a narrow range for challenge aerosols. The calibration models adjusted for particle CRI and density account for non-linearity in the OEM's mass concentrations estimates and demonstrated lower error. These results have significant implications for using PMS A003 sensor in high concentration environments, including indoor air quality monitoring, occupational/industrial exposure assessments, wildfire smoke, or near-source monitoring scenarios.

## 1. Introduction

The direct measurement of time- and size-resolved particle matter (PM) concentrations is essential to health-related applications, such as exposure assessments and air quality (AQ) studies, but are challenging to implement at fine spatial and temporal scales. Human exposure to PM is associated with multiple adverse health effects, including cardiovascular disease, cardiopulmonary disease, and lung cancer [1-7]. Estimates show that approximately 3% of cardiopulmonary and 5% of lung cancer deaths are attributed to exposures to $PM_{2.5}$ (particles less than 2.5 μm in diameter) globally. Particle deposition in the human respiratory tract and the resultant adverse health effects depend on the particles' size distribution [8, 9]. PM concentrations are known to vary significantly in space and time across community settings [10, 11]. Hence, time- and size-resolved PM measurements are more informative than traditional total PM weight measurements for assessing adverse health effects. The U.S. Environmental Protection Agency (EPA) has adopted the $PM_{10}$ and $PM_{2.5}$ criteria for monitoring air quality [12, 13]. However, the sparse spatial distribution of government monitoring sites makes fine spatial scale exposure assessment challenging [14]. Traditional PM instruments are large and expensive, thus have limited use in high spatial and temporal resolution mapping applications; these applications instead demand compact, low-cost sensors with reliable performance.

Low-cost particulate matter (PM) sensors find increasing use in various applications, including monitoring AQ in the outdoor [15-18] and indoor environment [19-21] by academic researchers and citizen scientists. The low-cost sensor networks have the potential to provide high spatial and temporal and resolution, identifying pollution sources and hotspots, which in turn can lead to the development of intervention strategies for exposure assessment and intervention strategies for susceptible individuals. Time-resolved exposure data from wearable monitors can be used to assess individual exposure in near real-time [22].

As low-cost sensors find applications in pollution monitoring, there is a need to characterize and calibrate their performance under various conditions, and especially in controlled environments with standardized test aerosols. Various studies have evaluated the performance of low-cost PM sensors in laboratory and field settings [23-30]; these reports show that low-cost sensors yield usable data when calibrated against research-grade reference instruments, although some drawbacks have also been reported. One common concern is that calibrations based on the number concentrations (not on the mass index) are not reported; however, it is essential since the mass concentration reported by the low-cost PM sensors is based on numerous assumptions. Second, there is a lack of information on the ability of low-cost sensors to assess particle size distributions, which is important for assessing health and environmental impacts. Furthermore, the particle number concentration in each size bin can be readily translated to the mass index with estimates of the particle CRI, shape, and density. Third, calibrations based on short-term field colocations with reference instruments typically cannot be used to generalize wide-ranging concentrations and the effect of varying particle sources. This is a concern because with improving air quality in the developed nations, where typical $PM_{2.5}$ levels are relatively low (<20 μg/m$^3$); however, PM concentration during wildfires [31] and in occupational settings [32, 33] often exceeds regulatory limits for short periods. Also, in developing countries with less strict regulations, the PM level associated with, e.g., traffic emissions [34], agricultural waste burning [35], indoor cooking [36] is significantly higher, but where field colocations for calibration studies can be conducted. Thus, evaluating low-cost PM sensors' performance under high and low loading conditions is necessary if the sensor were to be used in epidemiological studies and surveillance networks.

Optical particle sensors rely on elastic light scattering to measure time- and size-resolved PM concentrations; they are widely used in aerosol research, particularly when measuring particles in the 0.5 μm to 10 μm range. Aerosol photometers that measure the bulk light scatter of multiple particles simultaneously have limited success in laboratory studies [30]. Typical low-cost (<$100) particle monitors often yield unreliable number concentrations data [37], and PM mass estimation error can be as high as 1000% [38]. Also, low-cost sensor measurements may suffer from sensor-to-sensor variability due to a lack of quality control and differences between individual components [30, 37]. Sensor geometry can be optimized to reduce the effect of particle complex index of refraction (CRI). Researchers have addressed CRI sensitivity by designing optical particle sizers (OPSs) that measure scattered light at multiple different angles simultaneously [39] or by employing dual-wavelength techniques [40]. However, these solutions involve complex and expensive components not suitable for compact, low-cost devices. Optimizing the detector angle relative to the excitation beam can

reduce dependency on CRI [41]; however, this approach has not been translated to high volume fabrication.

Some commercially available low-cost sensors provide output in terms of total particle counts or particle mass concentrations, and some provide size-specific counts or mass concentrations. These quantities are not measured directly as an individual particle's scattering signatures (as in the single particle counters) but are estimated by the original equipment manufacturer's (OEM) proprietary algorithms. These algorithms have inherent limitations because particle scattering depends on the particles' composition, size, shape, and CRI [42]. A common workaround is to collect PM on a filter after or in parallel with the OPS measurements. The filters are analyzed to determine their average particle optical properties; these data are then used to correct the optical measurements after the fact.

Environmental conditions can affect sensor output; e.g., a non-linear response has been reported with increasing humidity (RH) [43-47]. High humidity (RH > 75%) creates challenges for particle instruments; e.g., significant variations were observed between different commercially available instruments, such as Nova PM sensor [43] and personal DataRAM [45]. In addition, the RH measurement approach could also affect the sensor output [43, 44], e.g., the RH measurement based on reference monitoring site rather than inside the sensor enclosure may be different due to the microenvironment and transient effects. The selection of reference instruments with different measuring principles may also influence the calibration of low-cost sensors. For example, the calibration of the Plantower PM sensor in Jayaratne et al., 2018 was based on the tapered element oscillating microbalance (TEOM), while Zusman et al., 2020 calibrated the same sensor against the beta attenuation monitor (BAM) and federal reference method (FRM) measurements [29, 44]. The integrated mass measurements cannot account for temporal particle size and concentration variation that can occur during the calibration experiment. The instruments that directly measure aerosol size and concentration in real-time can be a better fit for sensor calibration [30, 48]. The calibration against aerodynamic particle sizer (APS) or single optical particle counter instruments can potentially provide a more robust calibration for low-cost optical particle sensors. APS measures particle aerodynamic diameter based on the time-of-flight approach. For particles below the resolution threshold based on the aerodynamic principles ($d_p$ = 0.37-0.523 µm), APS uses an optical sensor reporting the total number of particles below the threshold. Correlating particle diameter measured low-cost sensor to aerodynamic diameter measured by an APS is relevant since the aerodynamic diameter determines particle deposition in the respiratory tract.

This study presents calibration for Plantower PMS A003 (Plantower, Beijing Ereach Technology Co., Ltd, China; referred to as PMS hereafter) sensors as a function of particle sizes and concentration, as well as the $PM_{2.5}$ and $PM_1$ indices. A standardized laboratory experimental protocol was developed to control the PM concentration, environmental conditions and to assess sensor-to-sensor reproducibility. The calibration is based on laboratory tests utilizing various polydisperse standard testing aerosols, including the Arizona Test Dust (ATD), two types of ceramic particles, and

NaCl particles. The PMS data from multiple sensors were compared against the APS for particle size range 0.5-10 micron and number density in the range of 0 - 1000 #/cc.

## 2. Material and Methods

### 2.1 Plantower PMS A003 and Sensor Test Platform

The low-cost sensor PMS A003 was evaluated. A sensor's photodiode positioned normal to the excitation beam measures the light scattered by the particles in the optical volume. The scattering light intensity is then converted to a voltage signal to estimate PM number density and mass concentration using a proprietary calibration algorithm. The PMS provides estimated particle counts in six size bins with the optical diameter in 0.3-10 µm (#/0.1L) range and mass concentration (µg/m$^3$) for $PM_1$, $PM_{2.5,}$ and $PM_{10}$. The mass concentrations are reported for two settings: "factory" and "atmospheric" conditions. The factory condition applies a correction factor (CF) of unity to the concentration measured, whereas the "atmospheric" condition is designed for use in the ambient environment.

Six PMS units were installed on a custom printed circuit board (PCB), which also included a Bosch BME680 temperature and relative humidity (RH) sensor (Fig. 1). All sensors were connected to an Arduino Nano microcontroller through a data selector (multiplexer NXP 74HC4051 breakout board, Sparkfun, Boulder, CO). The controller collects data from the six PMS sensors and an RH and temperature sensor simultaneously with the data acquisition rate of 1 Hz. The data reported in "factory" mode were used in the analysis.

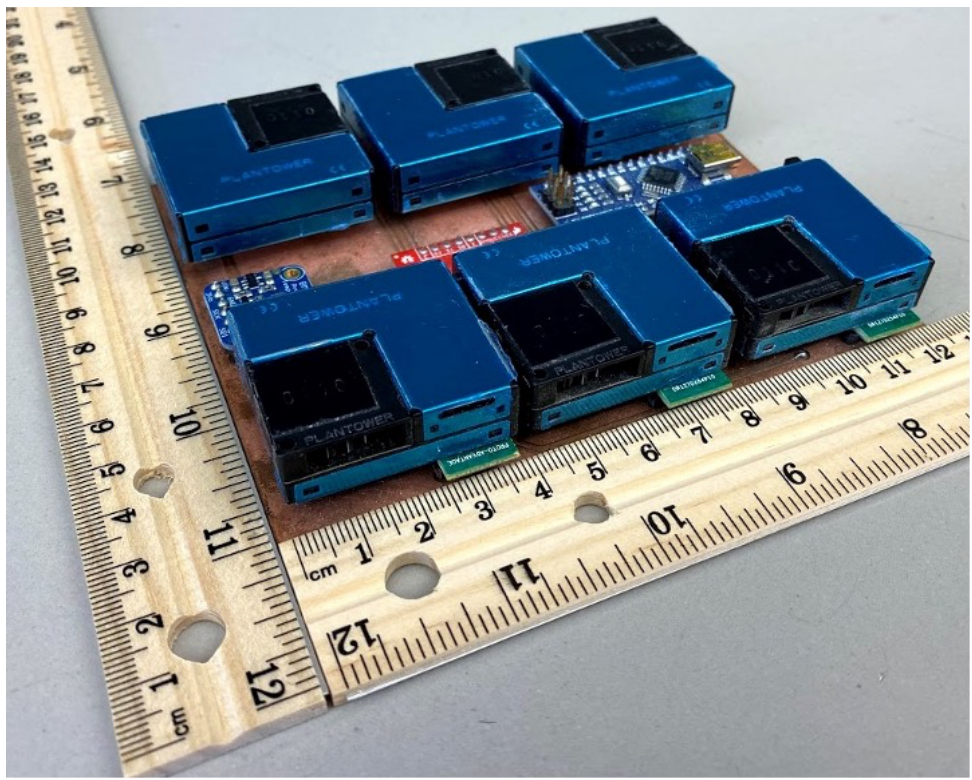

**Fig. 1. Photograph test platform, consisting of six PMS units mounted on the PCB, a temperature and humidity sensor, a multiplexer, and an Arduino microcontroller.**

The reference instrument of the experiment was the TSI Aerodynamic Particle Sizer (APS) Spectrometer 3321. APS measures real-time size-resolved particle counts with aerodynamic particle diameter ranging from 0.523 to 20 μm in 52 size bins using the time-of-flight principle. Particles smaller than 0.523 μm can be detected by the APS's optical sensor; however, their aerodynamic sizes were not resolved. Mass concentration is estimated by the instrument by assuming spherical particles and setting particle density values. During the experiments, APS continuously measured particle concentrations with a 5-second temporal resolution, the sampling inlet was placed in close proximity to the PMS sensors. The total number concentration of the aerosols in the experiments was maintained below 1000 #/cc ($10^5$ #/0.1L) to minimize the coincidence error in the APS measurement specified in the instrument's user manual.

**2.2 Aerosol Chamber Tests**

We tested four polydisperse aerosols: Arizona Test Dust (ATD) (Powder Technology Incorporated, Arden Hills, MN), polydisperse W210, and W410 ceramic particles (3M™, St. Paul, MN). NaCl particles were generated by nebulizing the aqueous solution of 10% wt [49]. The properties and typical size distributions of these aerosols are summarized in Table 1 and Fig A.3, respectively. The experiments were conducted in a custom-built aerosol chamber (0.56 m × 0.52 m × 0.42 m) (Fig. 2). A full description of the chamber can be found in ref [50]. The PMS sensor platform was placed in the well-mixed aerosol test chamber, elevated to the same height as the APS inlet. The APS sampled particle-laden air through static-dissipative tubing to eliminate electrostatic losses in the tubing. Particles were generated using a medical nebulizer (MADA Up-Mist Medication Nebulizer) [51]. During the experiments, the RH was controlled by nebulizing deionized water using a separate nebulizer or introducing dry filtered air; tests were conducted in the range of RH = 17% -

80%. Two mixing fans inside the chamber provided well-mixed conditions through the experiments; particle concentration was continuously monitored.

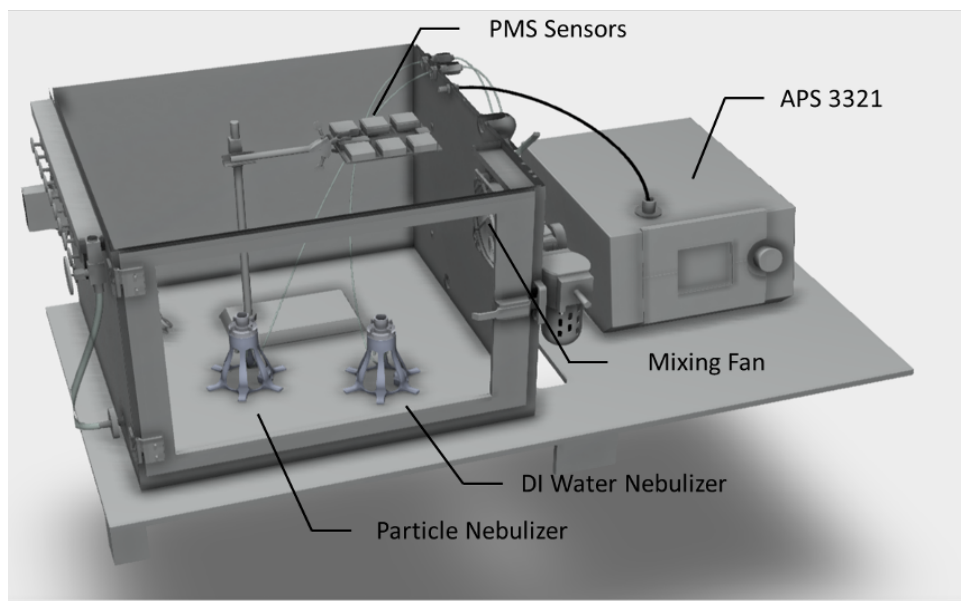

**Fig. 2. A 3D view of the experimental setup.**

**Table 1. Characteristics of the standard testing aerosols used in the study [52].**

| Aerosol | ATD | W210 | W410 | NaCl |
|---|---|---|---|---|
| Composition | $SiO_2$, $Al_2O_3$, $Fe_2O_3$, $Na_2O$ [a] | Alkali aluminosilicate ceramic | Alkali aluminosilicate ceramic | Sodium Chloride |
| Assumed density (g/cm³) | 2.5 – 2.7 [b] | 2.4 | 2.4 | 1.03 |
| CRI | 1.63 | 1.53 | 1.53 | 1.54 |

[a] Four major components were listed.

[b] For analysis purposes, 2.6 g/cm³ was used.

Definition of abbreviations: ATD = Arizona Test Dust

We controlled the aerosol generation rate by adjusting the compressed air flow rate to the nebulizer. The aerosol generation was stopped when the total number concentration (based on the APS count) reached 1000 #/cm$^3$. Then, the particle concentration was allowed to decay as the chamber was evacuated at a rate of 9.8 L/min; the make-up air entering the test is aspirated through a HEPA filter. The sensor array data and the APS data were acquired via two universal serial buses (USB) cables in real-time until the total number concentration from the APS reached 15 #/cm$^3$.

**2.3 Data Analysis and Modeling**

The collected data were checked for outliers, all measurements outside the measurement range of APS were removed. The number concentration reported by the APS was aggregated as summarized in Table 2 to match the cumulative number concentrations of the PMS. The 1-second PMS measurement and 5-second APS measurement were aggregated to obtain 1-minute averaged data for calibration. The smallest size bin of the APS (< 0.523 μm) was used as a reference for calibrating PMS size bin > 0.3 μm.

**Table 2. The PMS manufacturer's specified size bins and the corresponding reference APS size bins for calibration.**

| PMS size bin | Reference APS size bin |
|---|---|
| > 0.3 μm | aggregated from all size bins (< 0.523 μm and 0.542 – 19.81 μm) |
| > 0.5 μm | aggregated from size bins 0.542 – 19.81 μm |
| > 1 μm | aggregated from size bins 1.037 – 19.81 μm |
| > 2.5 μm | aggregated from size bins 2.642 – 19.81 μm |
| > 5 μm | aggregated from size bins 5.048 – 19.81 μm |
| > 10 μm | aggregated from size bins 10.37 – 19.81 μm |

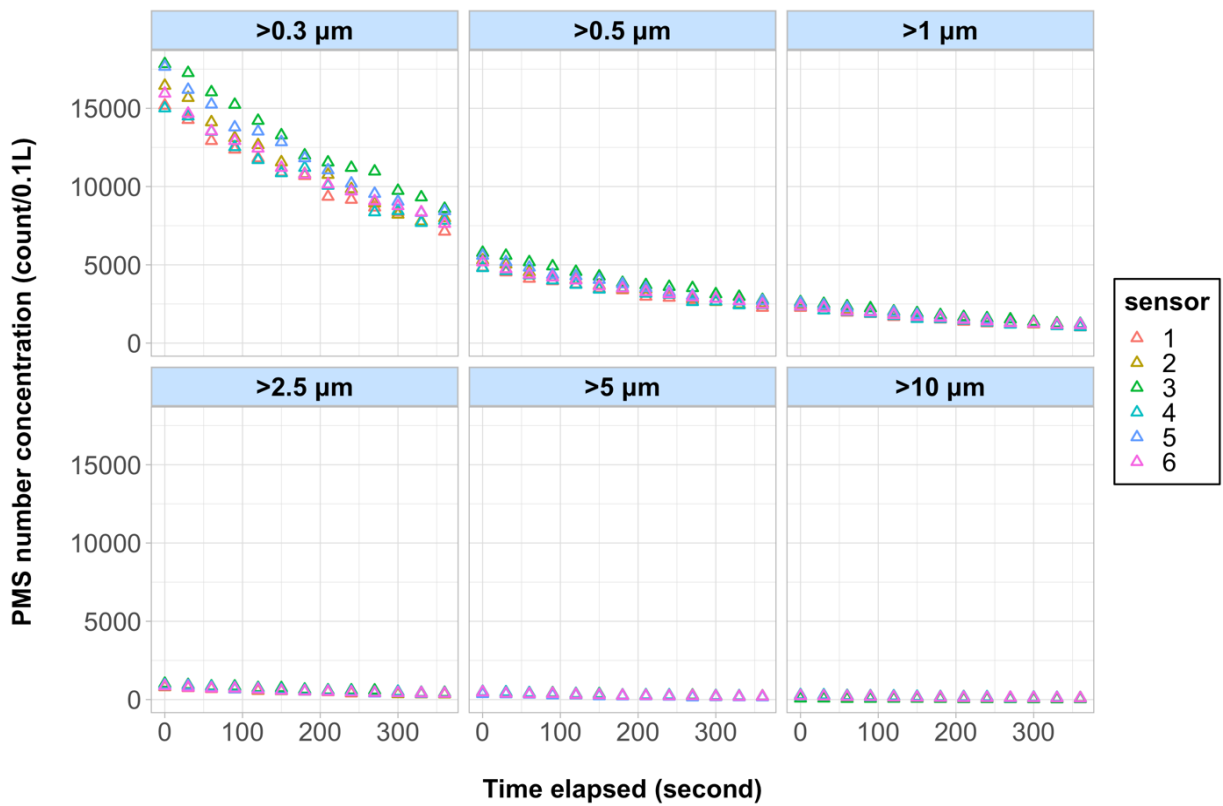

**Fig. 3. Time-series plots of the uncalibrated, 1-second number concentration measurement from the six PMS sensors during an experiment conducted under 30% relative humidity with W210 aerosols.**

Fig. 3 shows the data from 6 PMS sensors during the typical experimental run. In all experiments, the time-series of the uncalibrated concentration measurements from the six PMS sensors were consistent for all size bins (Pearson correlation coefficient (r) > 0.98) (Fig. A4 – Fig. A12). This allowed us to develop generalized models by fusing the readings from six sensors and then correlating the data against the APS measurement with matching time stamps for each size bin. For number concentration. Calibrations models of the following form were fit separately for number concentration data from the APS and PMS:

$$APS_t = \beta_0 + \beta_1 \, PMS_t + \varepsilon_t \quad (1)$$

where $APS_t$ is the number concentration for aggregated APS size bins at timestamp t; $PMS_t$ is the linear term of the PMS measurement (the number concentration of each PMS size bin) at timestamp t; $RH_t$ is the RH measurement of the Bosch BME680 sensor at timestamp t; $\beta_0$ and $\beta_1$ are regression coefficients; $\varepsilon_t$ is the residual. In addition to equation (1), other forms of linear models adjusted for relative humidity, particle density, and CRI were also evaluated (Table A.1). For calibration of mass concentration, models including quadratic terms of the PMS measurement were evaluated (Table A.1). Since the temperature variations for the tests were within ±2 °C, the temperature was not included as a variable in the models. Based on *a priori* condition that the PMS particle count and mass indices should be zero when the APS count is zero, the intercept ($\beta_0$) of the models was set to zero. The number of terms included in the optimal calibration model for each size bin was determined based on the Bayesian Information Criterion (BIC). The models with lower BIC were chosen as the optimal models. After identifying the optimal models using BIC, and estimating model coefficients, the model was then applied to the pre-calibrated 1-minute PMS measurement to produce the post-calibrated concentrations for model evaluation. Calibration performance was assessed using the normalized mean absolute error (NMAE), which was calculated using the following equation [53]:

$$\text{NMAE (\%)} = \frac{\text{Mean}(|C_{\text{PMS\_post-cal}} - C_{\text{APS}}|)}{\text{Mean}(C_{\text{APS}})} \qquad (5)$$

where $C_{\text{PMS\_post-cal}}$ is the post-calibrated 1-minute averaged PMS concentration, and $C_{\text{APS}}$ is the 1-minute averaged APS concentration. The linear models were fitted using the *lm* function in R. All the analyses were conducted using R version 3.6.3.

## 3. Results and Discussion

### 3.1 Test Conditions

During the experiments, the average temperature in the chamber was 24.8 °C (range: 23.2 to 26.6°C), the RH was varied in the range of 17.5 - 79.4%, all experiments were performed at atmospheric pressure conditions. The one-minute APS total number concentration averaged 237.9

#/cm$^3$ (range: 8.5 to 985.9 #/cm$^3$), the one-minute PM$_{2.5}$ measurement from the APS and PMS before calibration (6 sensors pooled together) averaged 106.0 µg/m$^3$ (range: 1.9 to 641.3 µg/m$^3$) and 51.5 µg/m$^3$ (range: 0 to 218.8 µg/m$^3$), respectively.

3.2 Particle Size Distribution

The particle size distribution of each test aerosol by APS is shown in Fig. A3. The NaCl particles (from the nebulized liquid solution) were the smallest among the test aerosols, with nearly all particles < 3 µm. The W410 mixture had slightly larger particles than W210, and it had had the same CRI as W210 [52]. The normalized particle size distribution of each test aerosol measured by the APS and the PMS was aggregated into six different size bins, as shown in Fig. 4. For W210 and W410, which had similar particle size distribution, the PMS indicated a minor difference in their size bin distributions. W410 had a slightly higher fraction of particles classified into larger size bins (2.5 – 5 µm and 5 –10 µm). For all aerosols and all tested concentrations, the PMS categorized some small fraction of particles into larger size bins (2.5–5 µm and 5–10 µm), which is not consistent with the APS measurements (Fig. 4). Overall, the PMS appeared to underestimate particle counts for the size bin 0.5–1 µm and 1–2.5 µm by a factor of 0.9 and 0.7, respectively. For larger size bins (2.5–5 µm and 5–10 µm), the PMS significantly overestimate the particle counts by a factor of 6 to 34 and 786 to 2097, respectively. The PMS also overestimated the counts in the 0.3–0.5 µm size range, although the APS range for comparison was larger (all particles <0.523 µm). PMS reflects the overall trends in particle size distribution; however, the absolute number concentration and particle sizing do not agree with the reference instrument. Thus, a better calibration based on a wider variety of aerosol types and environmental conditions is needed if PMS to be used for PM concentration assessment.

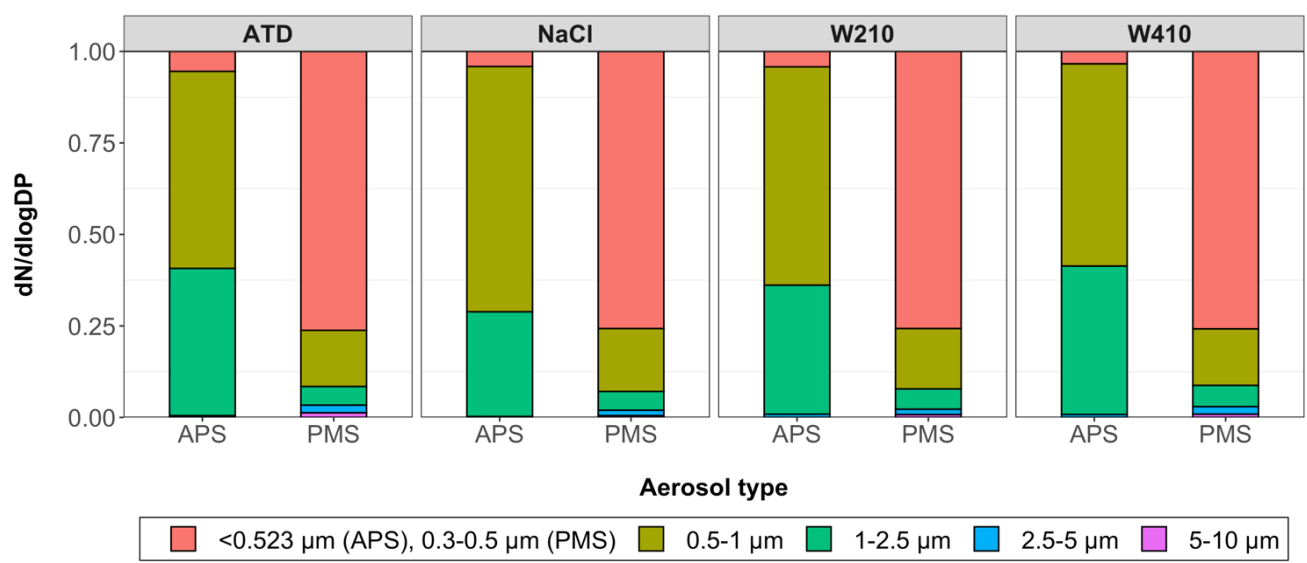

**Fig. 4. A comparison of the normalized size bin distribution measured by the PMS and APS before calibration. Typical data from one of the six PMS and APS for experiments with ATD, NaCl, W210, and W410 particles 15 minutes after the aerosols were introduced into the chamber. APS data from bins is aggregated to reflect the size ranges reported by PMS.**

**3.3 Model Fit**

Despite the apparent shift in size distribution shown by Fig. 4, a matrix of Pearson correlations between the PMS number concentrations (before calibration) and the APS reference number concentrations for different size ranges suggest that the matching size bins between the two instruments are well-correlated (Fig. A1). Notably, the PMS number concentration data correlated well with the APS for size bin up to 2.5 μm (r > 0.97). For measurement of size bin larger than 2.5 μm, the PMS exhibited moderate correlation with the APS (r <0.78). The worst correlation was observed for particles > 5 μm.

A similar Pearson correlation matrix comparing the mass concentrations measured by the PMS (before calibration) and the APS for different particle size ranges suggests a good correlation between matching sizes (Fig. A2). The $PM_1$, $PM_{2.5,}$ and $PM_{10}$ measurements by the PMS all exhibited high correlations with their corresponding sizes measured by the APS (r >0.90).

Because of the close correlations between corresponding APS and PMS size-specific measurements (Fig. A1 and Fig. A2), the sizes listed in Table 2 were chosen to develop calibration models for both PM number concentration and PM mass concentration. For example, APS size bin > 0.5 μm number concentrations was chosen as the reference (independent variable) for calibrating the PMS size bin > 0.5 μm number concentrations (dependent variable). Similarly, for mass concentrations, APS $PM_{2.5}$ mass concentration was chosen as the reference (independent variable) for calibrating the PMS $PM_{2.5}$ mass concentration (dependent variable).

After fitting a set of alternative calibration model forms to the APS and PMS number concentration data, the results show excellent $R^2$ and low NMAE for >0.3 μm, >0.5 μm, and >1 μm size bins when the full range of concentrations from 0 – 1000 #/cm$^3$ was used for fitting (Table 3 and

Table A.2). However, the model performance was worse for larger size ranges, i.e., >2.5 μm, >5 μm, and >10 μm size bins. Based on the previous findings on the impacts of relative humidity on optical particle sensor output and particle optical properties, the models adjusted for CRI and RH were considered in addition to the linear model, see Table 3. Interestingly, for most of the size ranges, the relatively simple linear model without the CRI dependent term performed nearly as well as the models with the additional parameters. Although based on the BIC, the optimal models (shown in bold in Table A.2) still tended to be those that included CRI and, in some cases, RH. The table also includes the performance of models based on the lower concentration data (data points with APS total number concentration <100 #/cm$^3$). These models do not perform as well even when applied to a lower range of particle concentrations.

**Table 3. Summary of the calibration model for number concentration, $R^2$, Bayesian information criterion (BIC) and the normalized mean absolute error (NMAE) of the calibration model.**

| Indices | Equation | Regression [a] | $R^2$ | BIC | NMAE |
|---|---|---|---|---|---|
| *Full concentration range (APS total number concentration 0 – 1000 #/ cm³) (n = 4,134)* | | | | | |
| >0.3 μm | Linear | y = 5.93 x | 0.99 | 78723 | 2.20% |
|  | Linear + CRI + RH | y = 6.00 x – 1090 CRI + 28.23 RH | 0.99 | 78567 | 2.06% |
| >0.5 μm | Linear | y = 14.17 x | 0.98 | 79716 | 2.92% |
|  | Linear + CRI + RH | y = 14.40 x - 1434 CRI + 40.68 RH | 0.98 | 79518 | 2.78% |
| >1 μm | Linear | y = 14.85 x | 0.96 | 76002 | 2.88% |
|  | Linear + CRI + RH | y = 14.98 x - 784.93 CRI + 26.40 RH | 0.97 | 75884 | 2.89% |
| >2.5 μm | Linear | y = 2.20 x | 0.66 | 62906 | 3.87% |
|  | Linear + CRI + RH | y = 2.42 x - 156.71 CRI + 3.38 RH | 0.68 | 62695 | 3.95% |
| >5 μm | Linear | y = 0.11 x | 0.31 | 38958 | 2.71% |
|  | Linear + CRI + RH | y = 0.14 x - 2.41 CRI - 0.06 RH | 0.33 | 38848 | 2.83% |
| >10 μm | Linear | y = 0.11 x | 0.70 | 8117 | 3.66% |
|  | Linear + CRI + RH | y = 0.14 x - 2.41 CRI - 0.06 RH | 0.71 | 8000 | 3.68% |
| *Lower concentration range (APS total number concentration < 100 #/ cm³) (n = 1,838)* | | | | | |
| >0.3 μm | Linear | y = 4.84 x | 0.97 | 30263 | 7.96% |
|  | Linear + CRI + RH | y = 4.94 x - 358.12 CRI + 13.68 RH | 0.97 | 30235 | 7.94% |
| >0.5 μm | Linear | y = 14.17 x | 0.93 | 30285 | 10.39% |
|  | Linear + CRI + RH | y = 10.83 x - 385.61 CRI + 18.91 RH | 0.94 | 30234 | 10.23% |
| >1 μm | Linear | y = 14.85 x | 0.92 | 26963 | 8.50% |
|  | Linear + CRI + RH | y = 11.10 x - 105.04 CRI + 10.59 RH | 0.93 | 26713 | 7.57% |
| >2.5 μm | Linear | y = 2.20 x | 0.6 | 14298 | 7.73% |
|  | Linear + CRI + RH | y = 0.45 x - 8.38 CRI + 0.48 RH | 0.66 | 14018 | 7.22% |
| >5 μm | Linear | y = 0.11 x | 0.35 | -2612 | 11.52% |
|  | Linear + CRI + RH | y = 0.003 x + 0.007 CRI + 0.002 RH | 0.44 | -2869 | 11.91% |
| >10 μm | Linear | y = 0.11 x | 0.19 | -5846 | 15.16% |
|  | Linear + CRI + RH | y = 0.002 x + 0.003 CRI + 0.0003 RH | 0.22 | -5920 | 17.91% |

[a] y: APS measurement; x: PMS measurement.

Definition of abbreviations: n = number of datapoints; CRI = complex index of refraction; RH = relative humidity; BIC = Bayesian information criteria; NMAE = normalized mean absolute error.

For fitting mass concentration data, an additional quadratic term was included, shown in Table 4. As with the number concentration models, restricting the mass concentration model to only lower concentrations resulted in worse performance vs. model based on the entire concentration range. The optimal models (shown in bold) included terms related to particle properties and environmental conditions (CRI, density, and RH). The improvements in NMAE for models with CRI, density, and RH terms compared to the relatively simple linear models without these added parameters tended to larger than observed for the number concentration models.

**Table 4. Summary of the calibration model for mass concentration, $R^2$, Bayesian information criterion (BIC), and the normalized mean absolute error (NMAE) of the calibration model.**

| Indices | Equation | Regression [a] | $R^2$ | BIC [b] | NMAE |
|---|---|---|---|---|---|
| *Full concentration range (APS total number concentration between 0 – 1000 #/ cm³) (n = 4,134)* ||||||
| $PM_1$ | Linear | $y = 1.06\ x$ | 0.96 | 25852 | 3.11% |
| | Polynomial | $y = 0.76\ x + 0.007\ x^2$ | 0.97 | 24480 | 2.41% |
| | Linear + CRI + density | $y = 1.13\ x + 13.88\ CRI - 10.13\ density$ | 0.97 | 24181 | 2.84% |
| | Polynomial + CRI + density | $y = 0.83\ x + 0.01 x^2 + 14.44\ CRI - 9.58\ density$ | 0.98 | 23432 | 2.33% |
| $PM_{2.5}$ | Linear | $y = 2.29\ x$ | 0.94 | 42435 | 4.53% |
| | Polynomial | $y = 1.55\ x + 0.006\ x^2$ | 0.96 | 41341 | 3.41% |
| | Linear + CRI + RH | $y = 2.51\ x - 23.27\ CRI + 0.36\ RH$ | 0.95 | 41565 | 4.07% |
| | Polynomial + CRI + RH | $y = 1.80\ x + 0.004\ x^2 - 15.55\ CRI + 0.42\ RH$ | 0.96 | 41152 | 3.44% |
| $PM_{10}$ | Linear | $y = 1.53\ x$ | 0.85 | 49963 | 3.56% |
| | Polynomial | $y = 0.72\ x - 0.003\ x^2$ | 0.88 | 48959 | 2.61% |
| | Linear + CRI + RH | $y = 1.69\ x - 39.14\ CRI + 0.56\ RH$ | 0.87 | 49544 | 3.31% |
| | Polynomial + CRI + RH | $y = 0.73\ x + 0.003\ x^2 - 17.94\ CRI + 0.75\ RH$ | 0.88 | 48931 | 2.61% |
| *Lower concentration range (APS total number concentration < 100 #/ cm³) (n = 1,838)* ||||||
| $PM_1$ | Linear | $y = 0.72\ x$ | 0.90 | 6211 | 10.10% |
| | Polynomial | $y = 0.91\ x - 0.02\ x^2$ | 0.91 | 6053 | 9.23% |
| | Linear + CRI + density | $y = 0.57\ x + 4.80\ CRI - 2.68\ density$ | 0.93 | 5746 | 8.16% |
| | Polynomial + CRI + density | $y = 0.80\ x - 0.02\ x^2 + 4.93\ CRI - 2.95\ density$ | 0.93 | 5694 | 8.08% |
| $PM_{2.5}$ | Linear | $y = 1.10\ x$ | 0.91 | 11170 | 9.14% |
| | Polynomial | $y = 1.34\ x - 0.01\ x^2$ | 0.91 | 11087 | 8.80% |
| | Linear + CRI + RH | $y = 0.97\ x - 1.87\ CRI + 0.16\ RH$ | 0.92 | 10890 | 8.01% |
| | Polynomial + CRI + RH | $y = 1.14\ x - 0.006\ x^2 - 2.43\ CRI + 0.17\ RH$ | 0.92 | 10885 | 7.97% |
| $PM_{10}$ | Linear | $y = 0.63\ x$ | 0.89 | 11878 | 9.30% |
| | Polynomial | $y = 0.86\ x - 0.01\ x^2$ | 0.90 | 11686 | 8.53% |
| | Linear + CRI + RH | $y = 0.54\ x - 2.15\ CRI + 0.20\ RH$ | 0.91 | 11627 | 8.01% |
| | Polynomial + CRI + RH | $y = 0.78\ x - 0.004\ x^2 - 3.57\ CRI + 0.21\ RH$ | 0.91 | 11461 | 7.75% |

[a] y: APS measurement; x: PMS measurement

Definition of abbreviations: n = number of datapoints; CRI = complex index of refraction; RH = relative humidity; BIC = Bayesian information criteria; NMAE = normalized mean absolute error.

Fig. 5 shows a comparison between the pre-calibrated and post-calibrated PMS and APS particle number densities for full and lower concentration ranges. The pre-calibrated (OEM) number concentration vs. APS exhibits a linear trend over the entire range for all aerosols; however, the PMS underestimates the number of particles. The calibration significantly improves the agreement demonstrating the importance of calibration and the accuracy gains from applying calibrations. The simple linear relationship shows excellent agreement over the entire range of particle concentration and properties, the fitting parameter are shown in Table 3.

For mass concentration (Fig. 6), the PMS does not increase linearly compared to APS estimates, especially at higher concentrations. We do not have a satisfactory explanation for the non-linear trend when using the OEM calibration. Also, we observed a notable discrepancy in the PMS and APS relationship between ATD and other test aerosols, which may be related to a wide range of particle CRI in ATD; see Table 1. The graphical comparison is consistent with our results from Table 4 that shows lower NMAE for the models with a quadratic term. Overall, the mass concentration models adjusted for particle and environmental specific properties such as CRI, density, RH, as well as adjustment for non-linearity seem to be necessary.

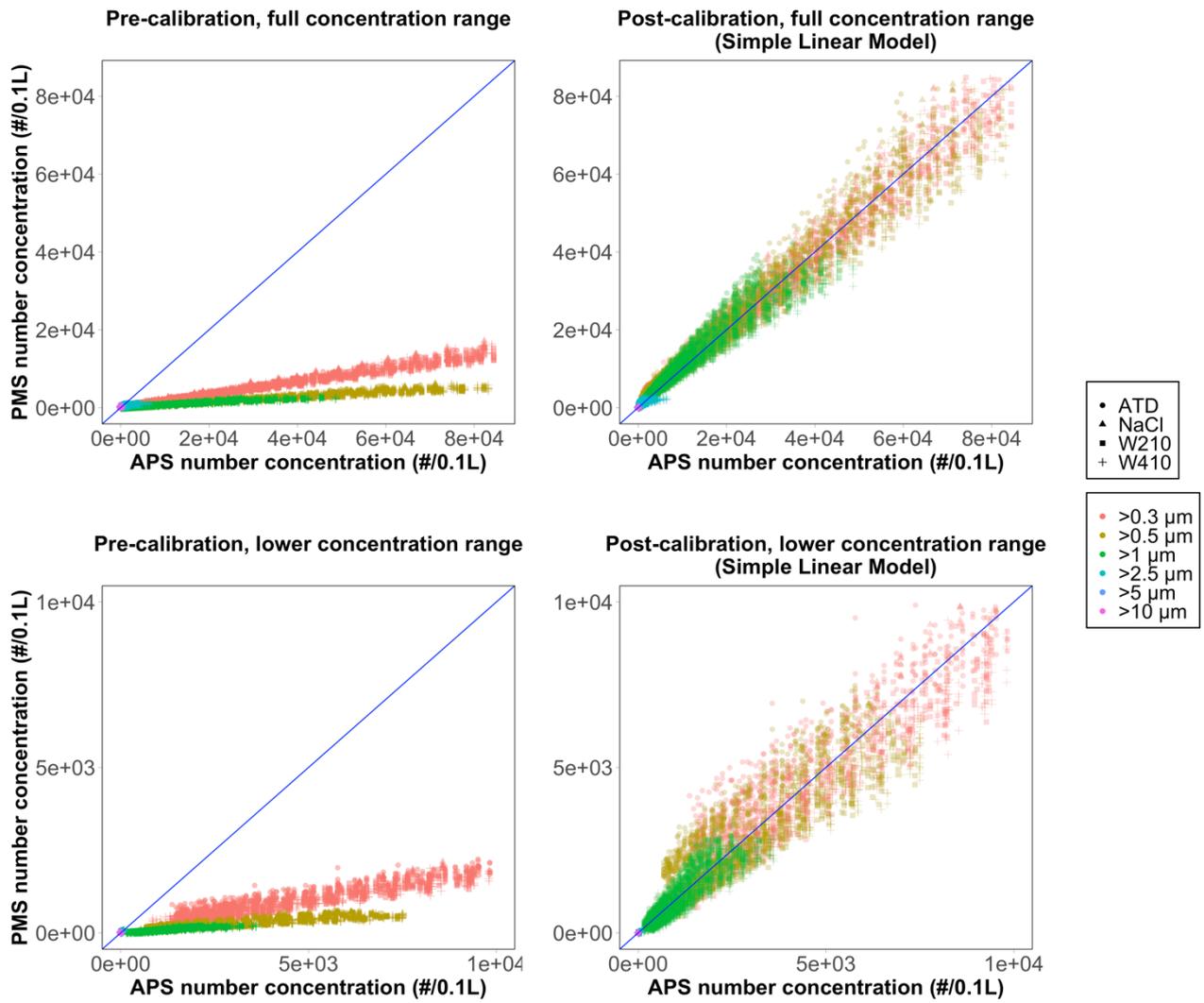

**Fig. 5. A comparison of the pre-calibrated and post-calibrated mass concentration by full and lower concentration range.** The blue line represents the 1:1 relationship between the PMS and APS concentration.

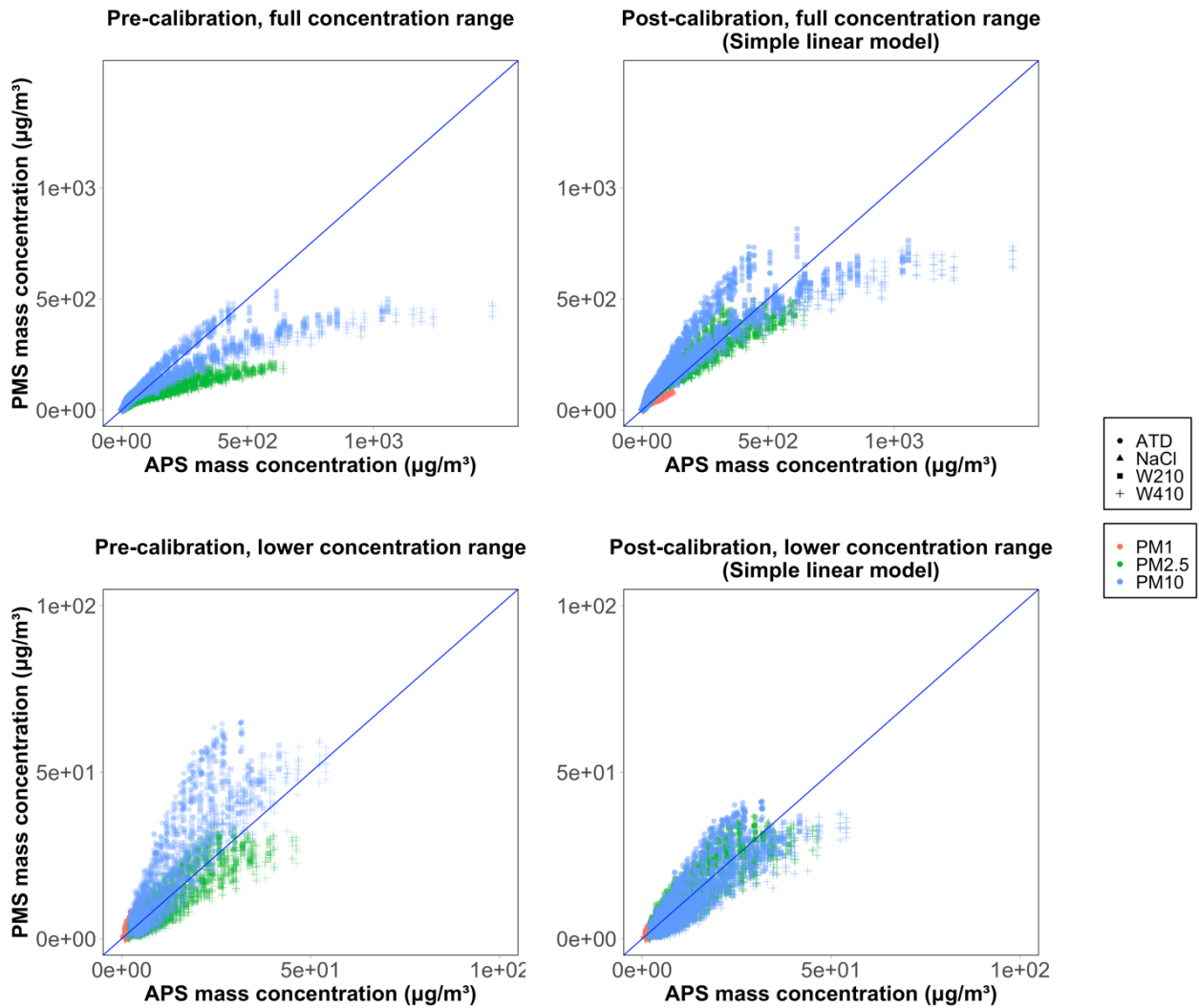

**Fig. 6. A comparison of the pre-calibrated and post-calibrated number concentration by full and lower concentration range.** The blue line represents the 1:1 relationship between the PMS and APS concentration.

## 4. Conclusions

In this study, we demonstrated not only the need for calibration of the low-cost PMS sensor but that, especially for mass concentration measurements, the manufacturer's provided estimates of mass concentration are inaccurate and respond non-linearly with increasing concentration. While we found that number concentrations may be reasonably calibrated using a simple linear calibration model (although BIC statistical considerations indicate that including particle and environmental conditions are technically superior calibration models), this was not the case for mass concentration calibrations. This study has major implications, especially for the use of these sensors in high concentration environments, which include many applications, including indoor air quality monitoring, occupational/industrial exposure assessments, wildfire smoke, or near-source monitoring scenarios. If number concentrations are of interest, our results suggest that although particle size distributions may be shifted compared to reference instrument measures, a linear calibration model without adjusting aerosol properties and RH, was able to correct the raw, uncalibrated low-cost PM sensor number concentration measurements for specific size bins with NMAE within 4.0% against the reference instrument. The models with the lowest BIC should be used with caution as the particle properties (CRI and density) used in model fitting were within a narrow range that based on four types of testing aerosols. However, if mass concentrations are of interest, our best calibration models accounted for non-linearity in the OEM's mass concentrations estimates, adjusted for particle CRI and density, and demonstrated lower error.

## 5. Acknowledgments

The authors wish to express special thanks to William Lin and Koustubh Muluk, the students at the University of Washington, for helping to run the aerosol chamber tests in this study.

## 6. Funding

This work was partially supported by the National Institute of Environmental Health Sciences [grant numbers 1R21ES024715 and 1R33ES024715]; and the National Institute of Biomedical Imaging and Bioengineering [grant number U01 EB021923].